\documentclass[aps,prb,preprintnumbers,amsmath,amssymb,superscriptaddress,twocolumn,10pt]{revtex4-2}
\usepackage{txfonts}
\usepackage{graphicx}
\usepackage{dcolumn}
\usepackage{bm}
\usepackage{array}
\usepackage[dvipsnames]{xcolor}
\usepackage[%
    colorlinks=true,
    pdfborder={0 0 0},
    linkcolor=blue
]{hyperref}  
\usepackage{chngpage}
\usepackage{appendix}
\usepackage{subfig}
\usepackage{booktabs}

\usepackage{caption}
\newcommand{\be}{\begin{equation}}
\newcommand{\ee}{\end{equation}}
\newcommand{\bea}{\begin{eqnarray}}
\newcommand{\eea}{\end{eqnarray}}

\begin{document}

\title{Origin of $\pi$-shifted three-dimensional charge density waves in kagome metal AV$_3$Sb$_5$}

\author{Heqiu Li}
	\affiliation{Department of Physics, University of Toronto, Toronto, Ontario M5S 1A7, Canada}
\author{Xiaoyu Liu}
\affiliation{Department of Physics, University of Toronto, Toronto, Ontario M5S 1A7, Canada}
	\affiliation{Department of Physics, University of Washington, Seattle, Washington 98195, USA}

\author{Yong Baek Kim}
\email{ybkim@physics.utoronto.ca}
	\affiliation{Department of Physics, University of Toronto, Toronto, Ontario M5S 1A7, Canada}
	\affiliation{School of Physics, Korea Institute for Advanced Study, Seoul 02455, Korea}

\author{Hae-Young Kee}
\email{hykee@physics.utoronto.ca}
	\affiliation{Department of Physics, University of Toronto, Toronto, Ontario M5S 1A7, Canada}
	\affiliation{Canadian Institute for Advanced Research, CIFAR Program in Quantum Materials, Toronto, Ontario M5G 1M1, Canada}

\date{\today}

\begin{abstract}

Understanding the nature of charge density wave (CDW) and superconductivity in kagome metal AV$_3$Sb$_5$ (A=Cs,Rb,K) is a recent subject of intensive study. Due to the presence of van Hove singularities, electron-electron interaction has been suggested to play an important role in the formation of such broken symmetry states. Recent experiments show that the CDW order is three-dimensional and it is staggered across different kagome layers. However, the experimental interpretation for the precise structure of CDW varies in terms of whether it is the star of David (SD), inverse star of David (ISD) or the alternation of the two among neighboring layers. In this work, we show that the origin of these distinct CDW orders can be understood in a unified picture by considering intra- and inter-layer electron-electron interactions as well as the coupling between electrons and lattice distortions. Utilizing an effective 9-band model with V $d$ orbitals and out-of-plane Sb $p$ orbitals, it is demonstrated that the repulsive electron-electron interaction favors charge bond order which induces either SD or ISD upon including lattice distortions. As the inter-layer interaction is introduced, $\pi$-shifted CDW develops with the staggered ordering along the $c$-axis. We also find that the phase with alternating SD and ISD can be stabilized as the ground state under strong inter-layer interaction. 


\end{abstract}


\maketitle

\section{Introduction }

Kagome materials provide a natural platform to study the interplay between many intriguing ingredients such as van-Hove singularities (vHS), flat bands, topology, Dirac cones, and geometric frustration. The family of vanadium-based kagome metal AV$_3$Sb$_5$ (A=Cs,Rb,K) has drawn great attention due to the recent discovery of various exotic phases in these materials~\cite{Jiang2022s,Neupert2022o,Wang2020s,Hu2022n,Oey2022,Christensen2022c,Zhu2022d,Stahl2022r,Wu2022d,Kang2022v,Li2021u,Wu2021u}. Superconductivity has been observed with $T_c\sim0.9-2.8K$~\cite{Ortiz2020,Ortiz2021,Chen2021,Chen2021b}, which coexists with charge density wave (CDW) with transition temperature $T_{CDW}\sim 80-100K$~\cite{Ortiz2019,Ortiz2020k,Shumiya2021,Ortiz2021s,Si2022,Song2022}. The CDW and SC phases observed in these materials are quite unconventional. Scanning tunnelling microscope (STM) experiments have shown that the CDW order induces lattice distortion in the vanadium kagome layers with enlarged $2\times 2$ periodicity~\cite{Jiang2021}, which is suggestive of the important role played by the vHS proximity to the Fermi level at momenta $M$. Rotational symmetry breaking is found in the CDW phase~\cite{Zhao2021,Li2022s}. Time-reversal symmetry breaking (TRSB) is observed in the CDW phase via muon spin relaxation ($\mu$SR) and STM experiments~\cite{Shumiya2021,Jiang2021,Mielke2022}, and a giant anomalous Hall effect~\cite{Shuo2020} is reported in the absence of magnetism~\cite{Kenney_2021}. However, Kerr effect measurements give contradicting results on TRSB~\cite{Xu2022n,Saykin2022,Hu2022s,Wang2023dk}.  Magnetoresistance measurements in ring structure sample in the superconductivity phase show oscillations corresponding to charge 4$e$ and 6$e$ flux quantization~\cite{Ge2022c}, indicating the possibility for novel superconductivity.

\begin{figure}
\includegraphics[width=3.2 in]{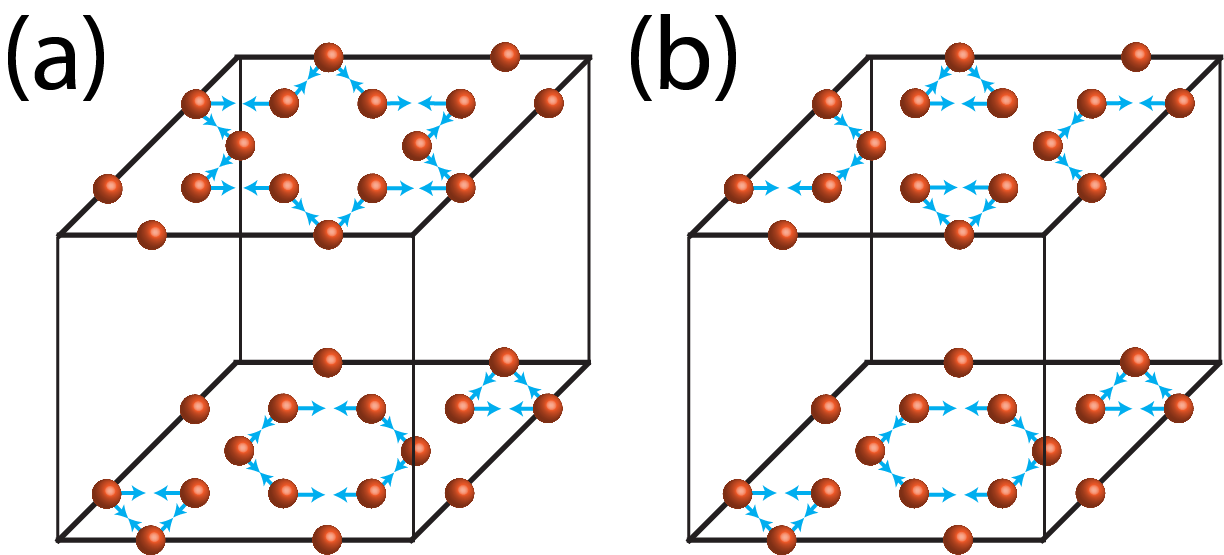}
\centering
\caption{ (a): Lattice distortion made of alternating SD and ISD patterns in neighboring kagome layers. Each site has displacement equal to the vector sum of the two arrows connected to it. (b): $\pi$-shifted ISD phase in which every kagome layer has ISD order but the pattern is shifted between neighboring layers.  }
\label{fig_order3d}
\end{figure}

The origin of superconductivity and the detailed structure of CDW still remain elusive. The CDW lattice distortion within the kagome planes compatible with $2\times 2$ periodicity is the star of David (SD) or inverse star of David (ISD), and the ISD is also called tri-hexagonal phase (TrH). These in-plane orders can also be stacked along the $c$-axis to give three-dimensional CDW order~\cite{Kang2022}. For example, in addition to the phase with uniform repetition of SD or ISD among different layers, there can be phases with alternating SD and ISD orders in neighboring layers as shown in Fig.\ref{fig_order3d}(a), or phases solely made of ISD(SD) but the CDW pattern is $\pi$-shifted between neighboring layers as in Fig.\ref{fig_order3d}(b), which is known as $\pi$-shifted ISD(SD) phase. Various experimental and theoretical techniques are applied to determine the structure of CDW~\cite{Nie2022,Li2021u,Stahl2022,Tan2021d,Liang2021c,Jiang2022e,Ritz2022,Ortiz2021sm,Hu2022t,Yang2022c,Luo2022s,Mu_2022,Luo2022n,Kato2022,Cho2021n,Denner2021,Park2021,Lin2021c,Christensen2021,Jeong2022,Rina2022,Zhou2022f,Dong2022,Ritz2022} and the results are diverse. The nuclear magnetic resonance (NMR) and nuclear quadrupolar resonance (NQR) experiments in Refs.\onlinecite{Frass2022} show that the CDW order is the $\pi$-shifted (staggered) ISD, whereas Ref.\onlinecite{Luo2022s} also performs NMR and NQR experiments but concludes that the CDW order is SD, and the combination of angle-resolved photoemission spectroscopy (ARPES) and density functional theory (DFT) in Ref.\onlinecite{Hu2022t} suggests the CDW order is the alternating SD and ISD order between neighboring layers. Therefore, a comprehensive understanding of the CDW order is highly desirable.

Given the close relationship between superconductivity and CDW in this family of kagome materials, identifying the nature of CDW can shed light on the study of superconductivity. For example, the symmetry and dimensionality of the CDW order parameter are crucial for modeling the normal state upon which the superconductivity is developed, and the mechanism that leads to the CDW order can provide valuable insight into the origin of superconductivity. The proximity of the van-Hove singularities at momentum $M$ to the Fermi level suggests that investigating the role played by electron interaction is crucial for understanding the nature of CDW and superconductivity.

\begin{figure}
\includegraphics[width=3.4 in]{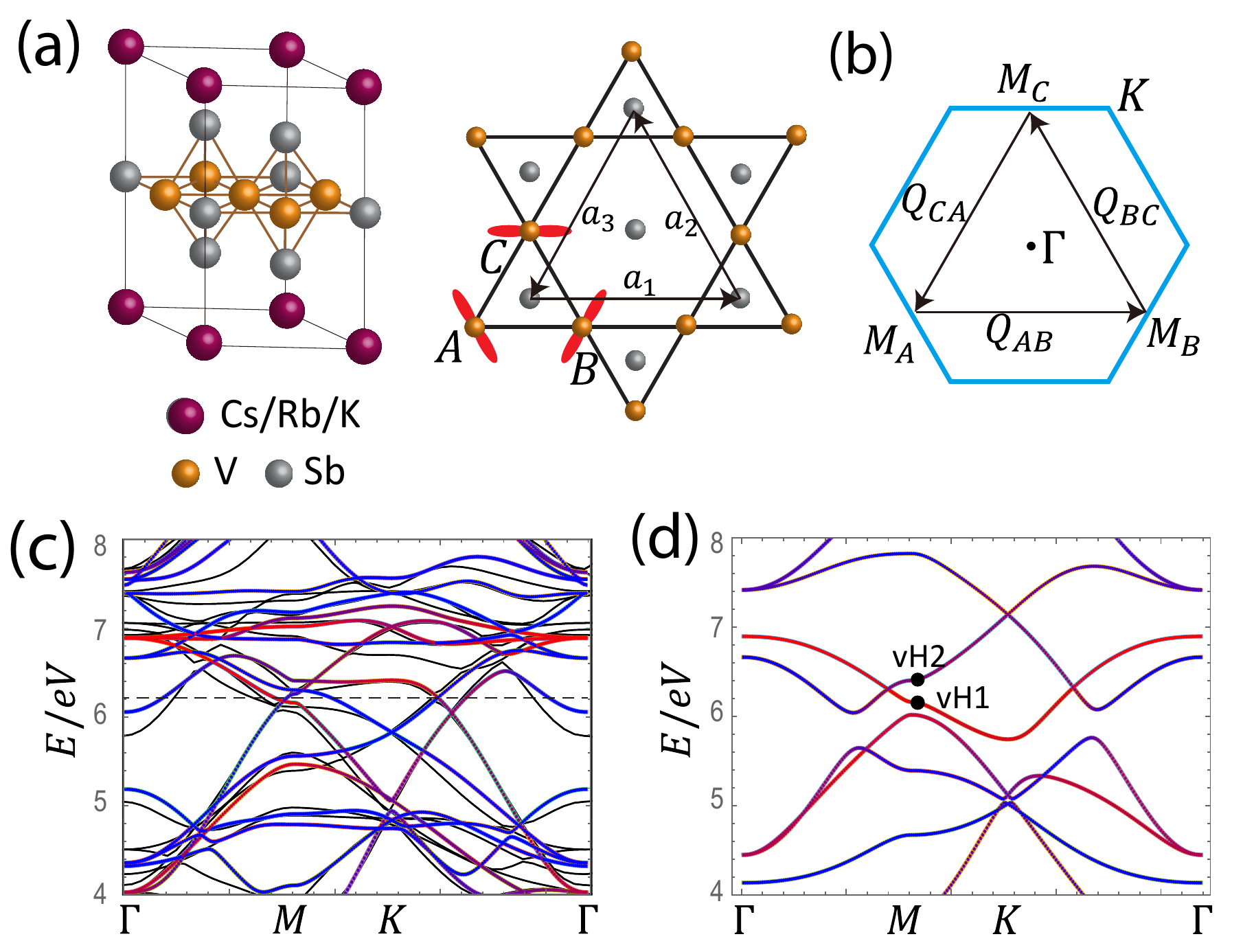}
\centering
\caption{ (a): Crystal structure of AV$_3$Sb$_5$ (A=Cs, Rb, K). The vanadium sites form a kagome lattice. The three red orbitals represent the $\tilde d$ orbitals obtained from superposition of $d_{xz},d_{yz}$ orbitals. (b): Brillouin zone of the lattice. (c): Band structure of the $30\times 30$ tight-binding model (red and blue) in comparison with the DFT band structure (black). The red color represent the weight of $d_{xz},d_{yz}$ orbitals in the wave function. (d): Band structure of the $9\times 9$ tight-binding model. vH1 and vH2 are van-Hove singular points.  }
\label{fig_crystal}
\end{figure}

\begin{figure*}[ht]
\centering
\includegraphics[width =  \linewidth]{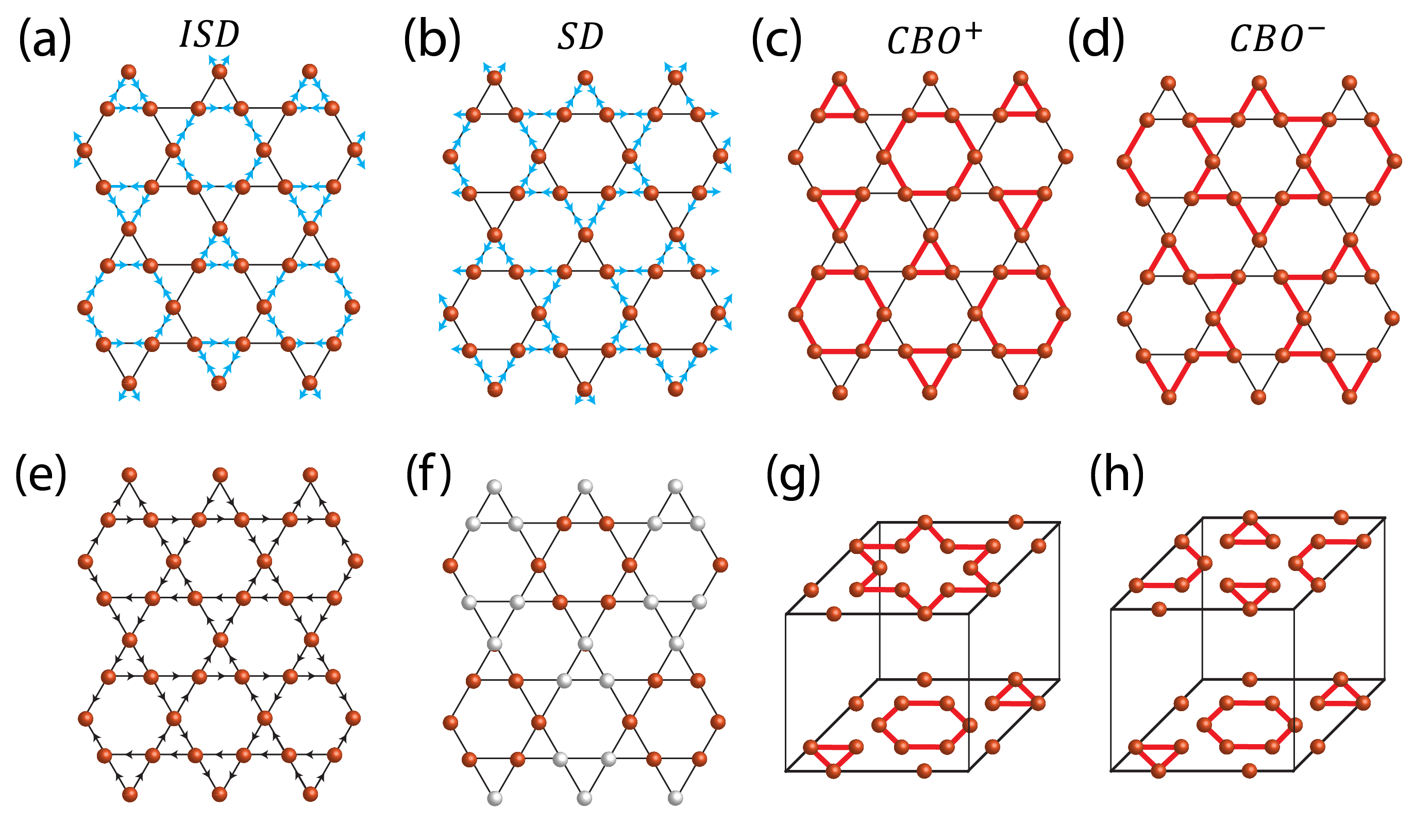}
\caption{ (a,b): Inverse star of David (ISD) and star of David (SD) patterns formed by lattice distortion. The displacement of each site is the vector sum of arrows connected to the site. (c,d): Electronic charge bond order CBO$^+$ and CBO$^-$ representing modulations of $\langle c^\dagger_{\mathbf r}c_{\mathbf r'}\rangle$ at nearest neighbor bonds. The bonds with positive modulation are marked in color. (e): Loop current order. The arrows represent the direction of current on each bond. (f): Charge density order. The on-site charge density is different at sites with different color. (g): Three-dimensional charge bond order with alternating CBO$^+$/CBO$^-$ order at neighboring kagome planes. (h): Three-dimensional charge bond order with $\pi$-shifted CBO$^+$ order.    }
\label{fig_order}
\end{figure*}

In this work, we show that the distinct CDW orders observed in experiments can be understood in a unified picture by considering electron interaction and the coupling between electron and lattice distortion. We construct a tight-binding model with both the $d$ orbitals from vanadium sites and $p$ orbitals from out-of-plane antimony sites using hopping parameters obtained from realistic DFT computation. Our mean-field computation shows that the intra-layer repulsive interaction stabilizes the electronic charge bond order (CBO) that involves three distinct $M$ points in the Brillouin zone. Furthermore, the inter-layer interaction along the $c$-axis favors non-uniform alignment of CBO among neighboring layers because it reduces the repulsion. Through the coupling between electron and lattice distortion, the CBO induces lattice distortion of either ISD or SD within each kagome layer. If inter-layer interaction is weak, it favors the $\pi$-shifted ISD or $\pi$-shifted SD order depending on the sign of electron-lattice coupling. If inter-layer interaction is strong, the phase with alternating ISD and SD in neighboring layers can be stabilized as the ground state because it can further lower the repulsive interaction. Therefore, this mechanism is a potential reason to explain the distinct three-dimensional CDW orders with enlarged periodicity $2\times 2\times 2$ observed in experiments.

\section{Tight-binding model with $d$ and $p$ orbitals}
\label{sec_TB}

The crystal structure of AV$_3$Sb$_5$ is characterized by the kagome planes of vanadium sites as shown in Fig.\ref{fig_crystal}(a), with one in-plane antimony Sb$_1$ and four out-of-plane antimony Sb$_2$ inside each unit cell. The bands near Fermi energy are made of superpositions of $d$ orbitals from vanadium sites and $p$ orbitals from antimony sites. We construct a tight-binding model by taking into account all these $d$ and $p$ orbitals, and determine the hopping amplitude among these orbitals by comparing with first-principle DFT bands for CsV$_3$Sb$_5$. This procedure yields a 30-band tight-binding model with five $d$ orbitals at each vanadium site and three $p$ orbitals at each antimony site. We choose the unit cell to contain three vanadium sites denoted by $A,B,C$ at an upper triangle as in Fig.\ref{fig_crystal}(a) and the antimony sites above and below the vanadium plane. This model includes up to second nearest neighbor in-plane V-V hopping and nearest neighbor V-Sb$_2$, V-Sb$_1$, Sb$_2$-Sb$_2$, Sb$_1$-Sb$_2$ hoppings. The band structure obtained from this model is shown in Fig.\ref{fig_crystal}(c), which agrees reasonably well with DFT computation. More details of this model are given in the appendix. The only inter-layer hopping is the hopping $t_{out}$ between out-of-plane Sb$_2$ atoms that are connecting different unit cells separated vertically along the $c$-axis. If $t_{out}=0$, then different layers are decoupled and the band dispersion does not depend on $k_z$. In reality there is a small hopping $t_{out}$ and the bands are weakly dispersing along the z direction.

To study the CDW phase with $2\times 2\times 2$ periodicity, we aim to simplify the model by focusing on the vHS at $M$ point of the Brillouin zone which mainly consists of $d_{xz}$ and $d_{yz}$ orbitals from vanadium sites. By choosing a suitable linear combination of $d_{xz}$ and $d_{yz}$ orbitals at each vanadium sites as the red orbitals in Fig.\ref{fig_crystal}(a), we find that the orbitals $\tilde d_\alpha$ ($\alpha=A,B,C$) contribute more than 80\% weight of the wave function near the vHS as shown in the appendix. Therefore, among the five $d$ orbitals at each vanadium sites we keep only one $\tilde d$ orbital. We also neglect the three $p$ orbitals at the in-plane antimony site Sb$_1$, because these orbitals mainly contribute to the $\Gamma$ point and have little contribution to the vHS. This reduces the 30-band model to a 15-band model with three $\tilde d$ orbitals and twelve $p$ orbitals at the four out-of-plane Sb$_2$ sites. 

We can further simplify this model by noticing that the $p$ orbitals at sites below and above the kagome plane can be superimposed to yield orbitals with positive or negative eigenvalues of the in-plane mirror symmetry $M_z$. The wave function at the vHS must be an eigenstate of $M_z$. Since the $\tilde d$ orbitals have negative eigenvalue of $M_z$, then among the twelve $p$ orbitals only the six orbitals with negative $M_z$ eigenvalue can contribute to the wave function at the vHS. Therefore we arrive at a 9-band model with three $\tilde d$ orbitals and six $p$ orbitals coming from linear combinations of out-of-plane $p$ orbitals with negative $M_z$ eigenvalues.

The band structure of this 9-band model is shown in Fig.\ref{fig_crystal}(d). The color of the bands represent the weight of wave function at vanadium $d$ orbitals and antimony $p$ orbitals. The red (blue) bands have more weight at $d$ ($p$) orbitals. The vHS can be seen at the $M$ points of the Brillouin zone, which are denoted by vH1 and vH2. The vH point vH1 is dominated by vanadium $\tilde d$ orbitals, while vH2 is a mixture of $\tilde d$ and $p$ orbitals. The dispersion away from the vH point deviates from the actual DFT band structure due to the fact that the model captures the band structure near the vHS while the features away from the vHS are neglected. 

{Compared with the commonly used three-band single-orbital model on kagome lattice, this model also includes the $p$ orbitals at the out-of-plane Sb$_2$ sites. These $p$ orbitals are important in modeling the CDW phenomena~\cite{Jeong2022,Tsirlin2022s,Ritz2022}. They are essential not only in reproducing the band structure close to DFT as in Fig.\ref{fig_crystal}(c), but also in capturing the nature of vHS vH1 and vH2. vH2 is a mixture of $p$ and $d$ orbitals, which will not exist if the $p$ orbitals of Sb$_2$ are not included. For the pristine type of vH point vH1 mostly made of $d$ orbitals, the $p$ orbitals are still important because the $p-d$ hopping modifies the effective hopping between $\tilde d$ orbitals to give it a correct sign. When modeling vH1 via the simple three-band model with only one orbital at each vanadium site, the hopping between nearest neighbor sites needs to be negative~\cite{Denner2021,Dong2022,Zhou2022f} to make the energy along $\Gamma M$ higher than $MK$ in the vicinity of vH1, as in Fig.\ref{fig_crystal}(d). However, our DFT computation shows the bare hopping between nearest neighbor $\tilde d$ orbitals is positive, and the hopping between $p$ and $d$ orbitals are comparable or even larger than the $d-d$ hopping (see the appendix for details). Therefore, the negative effective hopping between $\tilde d$ orbitals necessary for vH1 must come from the large coupling between $p$ and $d$ orbitals. This demonstrates the important role of $p$ orbitals from Sb$_2$ sites in modeling the van-Hove singularities.}

\section{Order parameter for different charge density waves}

Various of order parameters can lead to the enlarged $2\times 2$ periodicity within kagome planes. The order parameters can represent lattice distortion with the ordering wave vector at the three inequivalent $M$ points in the Brillouin zone. As shown in Fig.\ref{fig_order}(a,b), the kagome lattice of vanadium atoms can be distorted to form patterns of star of David (SD) or inverse star of David (ISD). The displacement of each site is the vector sum of the two arrows at that site. This distortion breaks the original translational symmetry and enlarges the unit cell. Without loss of generality we choose the unit cell to contain the three sites in the same upper triangle, i.e., the A,B,C sites in Fig.\ref{fig_crystal}(a). Label each unit cell by $\mathbf R_j$ and denote the strength of the arrows parallel to AB, BC, CA directions by $U_{AB}(j),U_{BC}(j),U_{CA}(j)$ respectively, then the ISD and SD phases with $2\times 2$ periodicity satisfy:
\begin{eqnarray}
U_{\alpha\beta}(j)&=&u_{\alpha\beta} \cos\left(\mathbf Q_{\alpha\beta} \mathbf R_j\right)
\label{uABC}
\end{eqnarray}
Here $\alpha\beta$ can take $AB,BC,CA$ and $Q_{AB},Q_{BC}$ and $Q_{CA}$ are vectors connecting different $M$ points in the Brillouin zone as shown in Fig.\ref{fig_crystal}(b). Each cosine factor in Eq.\eqref{uABC} is $\pm 1$. $u_{AB}=u_{BC}=u_{CA}>0$ represents an ISD phase and $u_{AB}=u_{BC}=u_{CA}<0$ represents and SD phase. Note that if the sign of one of the three $u$'s in an ISD (SD) phase is flipped, the phase will turn into SD (ISD) phase. If two of the three $u$'s are flipped, the type of the phase remains the same, but it is shifted by one lattice constant compared with the original phase. 

In addition to lattice distortion, electron-electron interaction can also lead to a broken translational symmetry with an enlarged unit cell. This includes the charge bond order (CBO), charge density order (CDO) and loop current order (LCO). Denote the electron creation operator of $\tilde d_\alpha$ orbital at vanadium site $\alpha=A,B,C$ of unit cell $j$ by $c^\dagger_{j,\alpha}$. The order parameter is given by:
\bea
\Delta_{\alpha\beta}=\frac{1}{N_c}\sum_j \left(\langle c^\dagger_{j,\alpha}c_{j,\beta}\rangle - \langle c^\dagger_{j,\alpha}c_{j-d_{\alpha\beta},\beta}\rangle \right)\cos(Q_{\alpha\beta}\mathbf R_j) .
\label{Dab}
\eea
Here $N_c$ is the number of unit cell, $\alpha\beta$ can be $AB,BC,CA$ and $d_{AB}=a_1,d_{BC}=a_2,d_{CA}=a_3$ respectively, where $a_1,a_2$ and $a_3$ are shown in Fig.\ref{fig_crystal}(a). The CBO and LCO phases are given by the real and imaginary part of $\Delta_{\alpha\beta}$. If $\Delta_{AB}=\Delta_{BC}=\Delta_{CA}=\Delta$ is real, then the corresponding CBO phases with $\Delta>0$ (CBO$^+$) and $\Delta<0$ (CBO$^-$) represent different modulations of $\langle c^\dagger_{j,\alpha}c_{j',\beta}\rangle$ at nearest neighbor bonds, which are shown in Fig.\ref{fig_order}(c) and (d) respectively. As we will show later, when the coupling between electron and lattice distortion is considered, CBO$^+$ will be stabilized as the ground state, which can induce ISD lattice distortion.

If $\Delta_{\alpha\beta}$ is imaginary, it represents the LCO with finite current $I\sim\langle i c^\dagger_{j,\alpha}c_{j',\beta}-i c^\dagger_{j',\beta}c_{j',\alpha}\rangle$ on nearest neighbor bonds. The LCO with $\Delta_{AB}=\Delta_{BC}=\Delta_{CA}=i|\Delta|$ is shown in Fig.\ref{fig_order}(e), where the arrows represent the direction of current on each bond. Contrary to the phases with lattice distortion or CBO, the LCO phase breaks time-reversal symmetry and can lead to the anomalous Hall effect.

There can also be CDO as shown in Fig.\ref{fig_order}(f) which modifies the electron density at each site. The order parameter is:
\be
\Delta_\alpha=\frac{1}{N_c}\sum_{j}\langle c^\dagger_{j,\alpha}c_{j,\alpha}\rangle \cos(M_\alpha\mathbf R_j)
\label{Dden2}
\ee
Here $\alpha=A,B,C$. Fig.\ref{fig_order}(f) corresponds to $\Delta_A=\Delta_B=\Delta_C>0$ with the origin chosen to be at the shaded upper triangle.

The order parameters in neighboring kagome layers can also be alternated to produce three-dimensional CDW order with $2\times 2\times 2$ periodicity. For example, among the CBO phases, if the signs of $\Delta_{AB},\Delta_{BC},\Delta_{CA}$ are alternating between neighboring layers as shown in Fig.\ref{fig_order}(g), it leads to a phase with alternating CBO$^+$ and CBO$^-$. This pattern is similar to Fig.\ref{fig_order3d}(a) but it represents electronic order rather than lattice distortion. If only $\Delta_{BC}$ and $\Delta_{CA}$ have alternating signs between layers and $\Delta_{AB}>0$ is fixed, the resulting phase is a $\pi$-shifted CBO$^+$ phase as shown in Fig.\ref{fig_order}(h). This pattern is similar to Fig.\ref{fig_order3d}(b) but without lattice distortion. If the sign of $\Delta_{\alpha\beta}$ alternates along z direction between layers, the ordering wave vector no longer connects $M$ points, but involves $L$ points in the Brillouin zone with $L=M+\frac{\pi}{c}\hat z$. Let $Q_{\alpha\beta L}\equiv Q_{\alpha\beta}+\frac{\pi}{c}\hat z$, then this new order parameter is defined as:
\bea
\Delta_{\alpha\beta L}=\frac{1}{N_c}\sum_j \left(\langle c^\dagger_{j,\alpha}c_{j,\beta}\rangle - \langle c^\dagger_{j,\alpha}c_{j-d_{\alpha\beta},\beta}\rangle \right)\cos(Q_{\alpha\beta L}\mathbf R_j) .
\eea
The phase with $\pi$-shifted CBO$^+$ in Fig.\ref{fig_order}(h) is represented by $(\Delta_{AB},\Delta_{BC},\Delta_{CA},\Delta_{ABL},\Delta_{BCL},\Delta_{CAL})\ =\ \Delta(1,0,0,0,1,1)$, which is also denoted by $MLL$ indicating it has one component at $M$ point and two components at $L$ point. The phase in Fig.\ref{fig_order}(g) is given by $(\Delta_{AB},\Delta_{BC},\Delta_{CA},\Delta_{ABL},\Delta_{BCL},\Delta_{CAL})\ =\ \Delta(0,0,0,1,1,1)$, which is denoted by $LLL$. Similar notations can also describe lattice distortion with alternating signs at different layers, this can be done by replacing $u_{\alpha\beta}$ and $Q_{\alpha\beta}$ in Eq.\eqref{uABC} by $u_{\alpha\beta L}$ and $Q_{\alpha\beta L}$ respectively.

\section{Full Hamiltonian: electron interaction and coupling to lattice distortion}

Various CDW orders can be induced by electron interaction. We consider systems at the filling of vH1 in Fig.\ref{fig_crystal}(d). The wave function at this vHS is dominated by $\tilde d$ orbitals which are superpositions of $d_{xz}$ and $d_{yz}$ orbitals at vanadium kagome sites as presented above. The Hamiltonian is written as:
\bea
H=H_0+H_V+H_{l}+H_c
\label{Htot}
\eea
The non-interacting part $H_0$ is the 9-band tight-binding model in Sec.\ref{sec_TB}. $H_V$ is the electron interaction between $\tilde d$ orbitals at kagome sites. Let $\mathbf r$ denote both the unit cell $j$ and sublattice $\alpha$ such that $c^\dagger_{\mathbf r}\equiv c^\dagger_{j,\alpha}$, $H_V$ is given by:
\be
H_V=V_d\sum_{\langle \mathbf r \mathbf r'\rangle} n_{\mathbf r}n_{\mathbf r'}+V_z\sum_{\mathbf r}n_{\mathbf r}n_{\mathbf r+c\hat z}
\ee
Here $n_{\mathbf r}=c^\dagger_{\mathbf r}c_{\mathbf r}$ and $c$ is the lattice constant along z direction. The $V_d$ term represent nearest neighbor (NN) interaction inside kagome plane, and the $V_z$ term represent interaction between sites at neighboring kagome layers that are on top of each other. $H_{l}$ is the elastic energy from lattice distortion in the kagome plane:
\be
H_l=\sum_{\langle \mathbf r \mathbf r'\rangle}\frac{1}{2}K_s(l_{\mathbf r \mathbf r'}-l_0)^2
\ee
Here $K_s$ is the spring constant and $l_0$ is the NN bond length without distortion. $\mathbf r$ and $\mathbf r'$ are restricted to be nearest neighbors inside the same kagome plane. $l_{\mathbf r \mathbf r'}-l_0$ is the change of distance between sites labeled by $\mathbf r$ and $\mathbf r'$ due to lattice distortion, which can be rewritten straightforwardly using $U_{\alpha\beta}(j)$ in Eq.\eqref{uABC}. For example, for the AB bond at an upper triangle in the unit cell labeled by $j$, the change of bond length is given by $2U_{AB}(j)+\frac{1}{2}U_{BC}(j)+\frac{1}{2}U_{CA}(j)$. $H_c$ is the coupling between electron and lattice distortion. Let the in-plane displacement of each site be $u_{\mathbf r}$ which is the vector sum of $U_{\alpha\beta}(j)$ connected to site $\mathbf r$, $H_c$ is given by:
\begin{eqnarray}
H_c&=&\lambda \sum_{\langle \mathbf r \mathbf r'\rangle}(l_{\mathbf r\mathbf r'}-l_0)(c^\dagger_{\mathbf r}c_{\mathbf r'}+c^\dagger_{\mathbf r'}c_{\mathbf r}) \nonumber\\
&&-\lambda_z\sum_\mathbf r (u_\mathbf r-u_{\mathbf r+c\hat z})^2 n_{\mathbf r}n_{\mathbf r+c\hat z}
\label{Hcoup}
\end{eqnarray}
The $\lambda$ term comes from the change of effective hopping strength due to the change of bond length $l$. If the hopping term is $-t(l)(c^\dagger_\mathbf r c_{\mathbf r'}+c^\dagger_{\mathbf r'} c_\mathbf r)$ with $t(l)>0$ and $\frac{dt}{dl}<0$, Taylor expansion leads to $-\frac{dt}{dl}(l_{\mathbf r\mathbf r'}-l_0)(c^\dagger_\mathbf r c_{\mathbf r'}+c^\dagger_{\mathbf r'} c_\mathbf r)$, which gives $\lambda=-\frac{dt}{dl}>0$. If $\lambda$ is positive, then a NN bond with $\langle c^\dagger_{\mathbf r}c_{\mathbf r'}+c^\dagger_{\mathbf r'}c_{\mathbf r}\rangle>0$ can induce a shrink of bond length, i.e., the CBO$^+$ (CBO$^-$) phase can induce ISD (SD) phase through coupling between electron and lattice distortion. In practice, the sign of $\lambda$ may depend on microscopic details. The $\lambda_z$ term comes from the fact that when $u_\mathbf r\ne u_{\mathbf r+c\hat z}$, the two sites with $V_z$ interaction are separated further, which reduces the $V_z$ interaction and leads to Eq.\eqref{Hcoup} with $\lambda_z>0$. Note that the displacement $u_\mathbf r$ and the change of bond length $l_{\mathbf r\mathbf r'}-l_0$ can be written in terms of $U_{\alpha\beta}(j)$ and $u_{\alpha\beta}$ in Eq.\eqref{uABC}. Therefore we will use $u_{\alpha\beta}$ as the order parameter to represent lattice displacement.

\section{Mean field analysis of charge density waves}

\subsection{Two-dimensional order parameters within kagome planes}
\label{sec_2D}

We use mean field theory to determine which order parameter is favored in the ground state. For simplicity we first study the effective two-dimensional system obtained by turning off the coupling between different kagome layers, which includes $V_z,\lambda_z$ terms and the inter-layer hopping $t_{out}$ between different Sb$_2$ sites in different unit cells. In this limit the band structure becomes two-dimensional and does not depend on $k_z$. 

The repulsive interaction $V_d$ at NN bonds can be decoupled by a Hubbard-Stratonovich (H-S) transformation. Define the Fourier transformation of electronic operator as $c^\dagger_{\mathbf k,\alpha}=\frac{1}{\sqrt{N_c}}\sum_j e^{i\mathbf k\mathbf R_j}c^\dagger_{j,\alpha}$. For bonds along $\alpha\beta$ direction ($\alpha\beta=AB,BC,CA$) at an upper-triangle of the kagome lattice, the interaction $H_V^u$ and its mean-field decoupling $H_V^{u,MF}$ are given by:
\begin{eqnarray}
&&H_V^u=V_d \sum_{j} c^\dagger_{j,\alpha} c_{j,\alpha} c^\dagger_{j,\beta} c_{j,\beta} = -\frac{V_d}{N_c}\sum_{\mathbf k\mathbf k'\mathbf q} c^\dagger_{\mathbf k+\mathbf q,\alpha}c_{\mathbf k,\beta}c^\dagger_{\mathbf k'-\mathbf q,\beta}c_{\mathbf k',\alpha} \nonumber\\
&&\ \ \ \ \ \ = -\frac{V_d}{N_c}\sum_{\mathbf k\mathbf k'} c^\dagger_{\mathbf k+\mathbf Q_{\alpha\beta},\alpha}c_{\mathbf k,\beta}c^\dagger_{\mathbf k'-\mathbf Q_{\alpha\beta},\beta}c_{\mathbf k',\alpha}, \nonumber\\
&&H_V^{u,MF}= -V_d\sum_{\mathbf k} \left(\Delta^u_{\alpha\beta}c^\dagger_{\mathbf k+\mathbf Q_{\alpha\beta},\beta}c_{\mathbf k,\alpha}+h.c. \right)+N_c V_d|\Delta^u_{\alpha\beta}|^2 .
\label{Vdmfu}
\end{eqnarray}
Here we only keep $\mathbf q=\mathbf Q_{\alpha\beta}$ in accordance with the $2\times 2$ CDW. We have introduced $\Delta^u_{\alpha\beta}=\frac{1}{N_c}\sum_{\mathbf k}\langle c^\dagger_{\mathbf k+\mathbf Q_{\alpha\beta},\alpha}c_{\mathbf k,\beta}\rangle$ in the H-S transformation. Since we have chosen the unit cell to contain three sites at upper triangles, for the bonds at down triangles the transformation of $V_d$ has an additional $\mathbf k$-dependent phase. The corresponding interaction $H_V^d$ and its mean-field decoupling $H_V^{d,MF}$ are:
\begin{eqnarray}
&&H_V^d=V_d \sum_{j} c^\dagger_{j,\alpha} c_{j,\alpha} c^\dagger_{j-d_{\alpha\beta},\beta} c_{j-d_{\alpha\beta},\beta}, \nonumber\\
&&H_V^{d,MF}= -V_d\sum_{\mathbf k} \left(\Delta^d_{\alpha\beta}e^{i\mathbf k d_{\alpha\beta}} c^\dagger_{\mathbf k+\mathbf Q_{\alpha\beta},\beta}c_{\mathbf k,\alpha}+h.c. \right)+N_c V_d|\Delta^d_{\alpha\beta}|^2, \nonumber\\
\label{Vdmfd}
\end{eqnarray}
where $\Delta^d_{\alpha\beta}=\frac{1}{N_c}\sum_{\mathbf k}\langle c^\dagger_{\mathbf k+\mathbf Q_{\alpha\beta},\alpha}c_{\mathbf k,\beta}\rangle e^{i\mathbf k d_{\alpha\beta}}$. The mean field Hamiltonian $H_{MF}$ is obtained by replacing the $H_V$ term in Eq.\eqref{Htot} by Eq.\eqref{Vdmfu} and \eqref{Vdmfd}, and $\alpha\beta$ is summed over AB,BC and CA. The simultaneous presence of $Q_{AB},Q_{BC},Q_{CA}$ gives rise to the $2\times 2$ periodicity and shrinks the Brillouin zone to one-quarter of its original size. The free energy $F$ can be computed from the eigenvalues of $H_{MF}$. When lattice distortion described by $u_{\alpha\beta}$ is also considered, $F$ is a function of $\Delta^u_{\alpha\beta},\Delta^d_{\alpha\beta}$ and $u_{\alpha\beta}$ with $\alpha\beta=AB,BC,CA$. Minimization of $F$ by changing these variables gives the mean field solution for these order parameters.

\begin{figure}
\includegraphics[width=3.4 in]{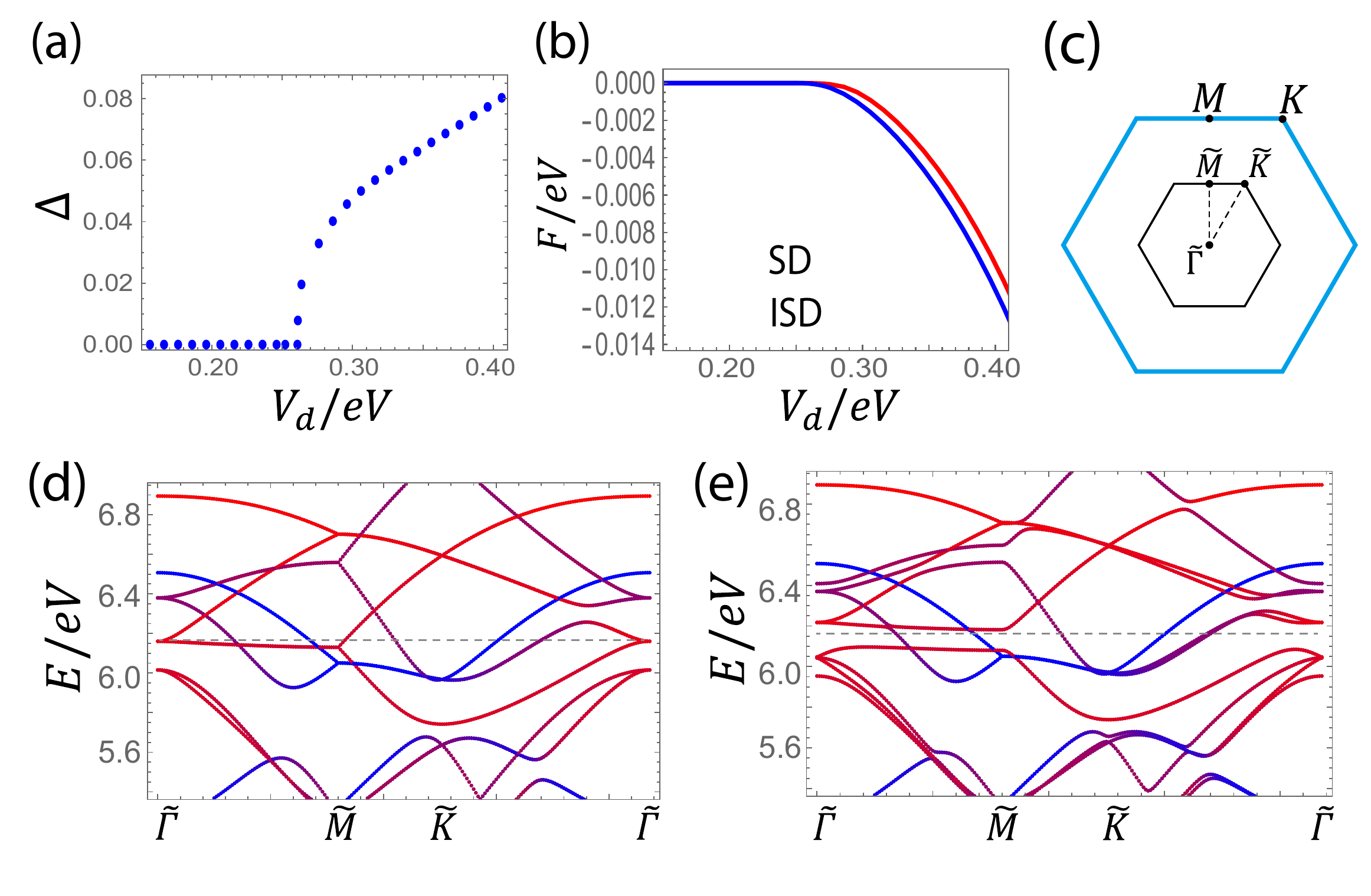}
\caption{ (a): The order parameter $\Delta$ of ISD phase as a function of interaction $V_d$ in the limit of decoupled layers. (b): Free energy $F$ as a function of $V_d$ for ISD and SD phases. (c): The reduced (original) Brillouin zone is shown as the black (blue) hexagon. (d): Band structure in reduced Brillouin zone without the development of any order parameter. The dashed line represents the van-Hove point vH1. The red (blue) bands are dominated by $d$ ($p$) orbitals. (e): Band structure in reduced Brillouin zone with ISD order $\Delta=0.06,u/l_0=0.01$. The order parameter opens a gap at van-Hove point but the system remains gapless.      }
\label{fig_Delta2d}
\end{figure}

We find a finite mean field solution corresponding to the global minimum of free energy when $V_d$ increases. The solution shows both $\Delta^u_{\alpha\beta}$ and $\Delta^d_{\alpha\beta}$ are real numbers, and they always have opposite sign $\Delta^u_{\alpha\beta}=-\Delta^d_{\alpha\beta}$. Therefore the solution can be characterized by their difference $\Delta_{\alpha\beta}=\Delta^u_{\alpha\beta}-\Delta^d_{\alpha\beta}$. Note that the $\Delta_{\alpha\beta}$ here is the same as the order parameter in Eq.\eqref{Dab}. The mean field solution also preserves the threefold rotation symmetry, with $\Delta_{AB}=\Delta_{BC}=\Delta_{CA}\equiv\Delta$ and $u_{AB}=u_{BC}=u_{CA}\equiv u$. {When the electron-lattice coupling $\lambda$ is positive, the solution has $\Delta>0$ and $u>0$, which represents coexisting phases of electronic charge bond order CBO$^+$ and lattice distortion with ISD pattern. If the sign of $\lambda$ is flipped $\lambda<0$, the solution shows that $\Delta$ remains positive but $u$ becomes negative, representing SD type of lattice distortion. Without loss of generality, from now on we focus on the case with $\lambda>0$. $\Delta$ as a function of $V_d$ is shown in Fig.\ref{fig_Delta2d}(a) with the parameter choice $T=92K,\lambda =0.5eV/l_0,K_s =1.4eV/l_0^2$ where $T$ is the temperature. The phase with CBO$^-$ and SD lattice distortion ($\Delta<0,u<0$) is a local minimum of free energy, but not the global minimum. The free energy for phases with ISD ($\Delta>0,u>0$) and SD ($\Delta<0,u<0$) order are shown in Fig.\ref{fig_Delta2d}(b), which shows ISD phase has lower free energy. The free energy is not an even function of order parameters as seen by the energy difference between SD and ISD phases, hence the phase transition that leads to a finite $\Delta$ is different from a usual second-order transition.}

{The order parameters in the mean field solutions are real numbers. We find that the solution with imaginary $\Delta$ corresponding to the LCO can be local minimum of free energy when interaction is large, but it has higher energy than the real CBO order. Hence the LCO is not the ground state in our computation. This is consistent with Ref.~\onlinecite{Dong2022} which shows that complex charge bond order corresponding to loop current phases only occurs when there is large second-nearest neighbor interaction.}

The band structure in the reduced Brillouin zone (Fig.\ref{fig_Delta2d}(c)) before and after the development of the ISD phase is shown in Fig.\ref{fig_Delta2d}(d,e). When $\Delta=u=0$, the band structure is in Fig.\ref{fig_Delta2d}(d). The dashed line represents vH1 and the red (blue) bands are dominated by $d$ ($p$) orbitals. At finite interaction $V_d=0.32eV$, ISD order is developed with $\Delta=0.06,u/l_0=0.01$ and the band structure is in Fig.\ref{fig_Delta2d}(e). The ISD order opens a gap near the vHS which is dominated by $d$ orbitals at vanadium sites, but due to the presence of other $p$ orbitals at antimony sites the system remains a metal after the charge order is developed.

\subsection{Three-dimensional structure of order parameters}
 
\begin{figure}
\includegraphics[width=3.4 in]{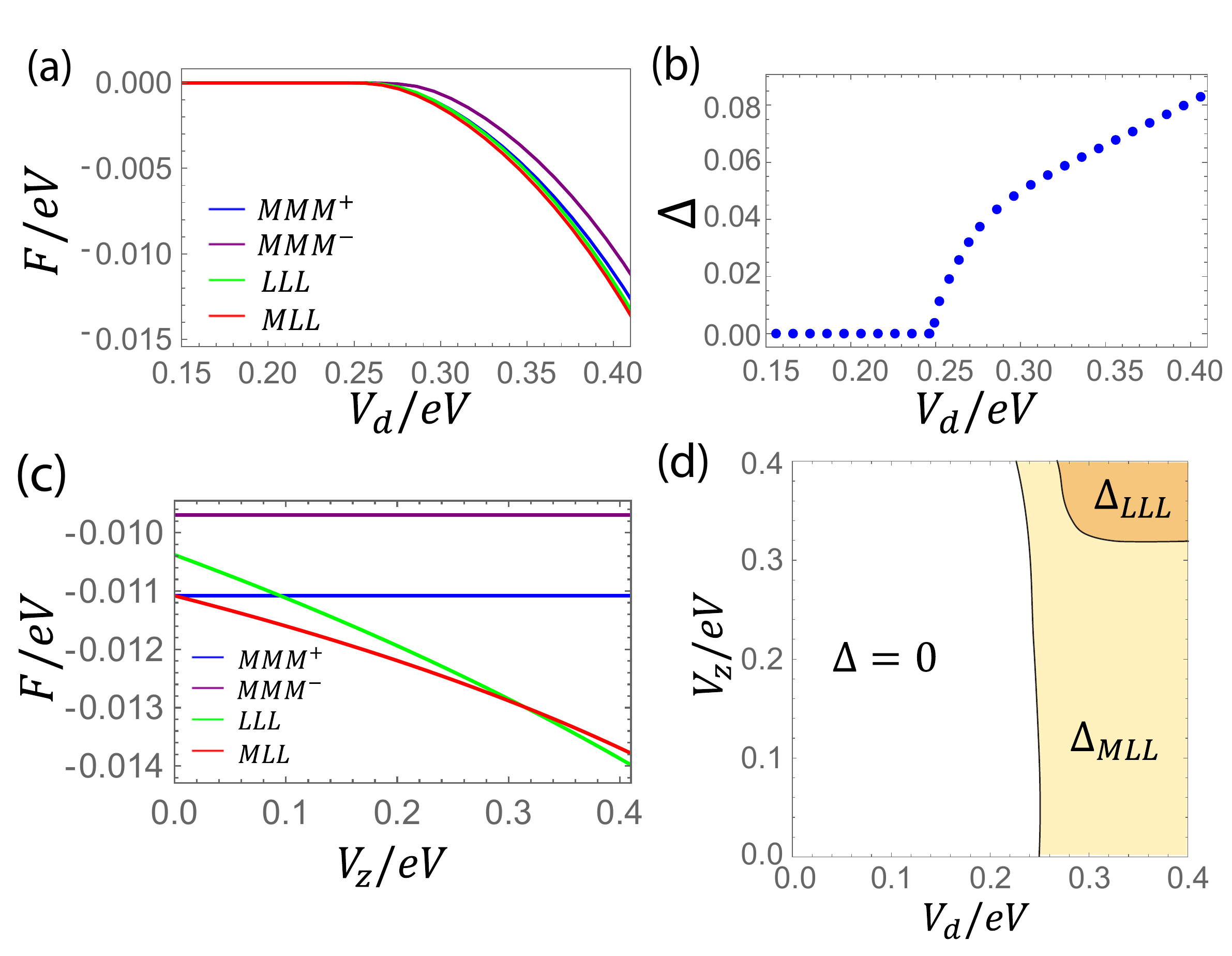}
\caption{ (a): Free energy $F$ as a function of interaction $V_d$ for different phases. Here we take $V_z=0.4 V_d$. (b): Order parameters of the shifted ISD phase as a function of $V_d$. (c): Free energy $F$ as a function of $V_z$ for different phases at fixed $V_d=0.4eV$. (d): Phase diagram for parameters $V_z$ and $V_d$. At small $V_z$ the system develops the shifted ISD order $\Delta_{MLL}$ as $V_d$ increases. If $V_z$ is large the phase $\Delta_{LLL}$ with alternating ISD and SD order at neighboring layers can be the ground state.  }
\label{fig_Delta3d}
\end{figure}

Now we turn on finite interaction $V_z,\lambda_z$ along z direction and recover the realistic inter-layer hopping $t_{out}$ to study the three-dimensional order parameters. The $V_z$ interaction can be rewritten as $V_z n_{\mathbf r}n_{\mathbf r+c\hat z}=\frac{V_z}{4}(n_{\mathbf r}+n_{\mathbf r+c\hat z})^2-\frac{V_z}{4}(n_{\mathbf r}-n_{\mathbf r+c\hat z})^2$, therefore it favors a phase with $n_{\mathbf r}\ne n_{\mathbf r+c\hat z}$ which will lower the $V_z$ repulsion, i.e., a phase with $\langle n_{\mathbf Q,\alpha}\rangle\ne 0$ and $Q_z=\frac{\pi}{c}$. To be consistent with the $2\times 2$ periodicity in the kagome plane, a natural choice of the order parameter is the three-dimensional generalization of Eq.\eqref{Dden2}:
\begin{eqnarray}
\Delta_{\alpha L}&=&\frac{1}{N_c}\sum_{j}\langle c^\dagger_{j,\alpha}c_{j,\alpha}\rangle \cos(M_{\alpha L}\mathbf R_j)\nonumber\\
&=&\frac{1}{N_c}\sum_{\mathbf k}\langle c^\dagger_{\mathbf k+M_{\alpha L},\alpha}c_{\mathbf k,\alpha}\rangle,
\end{eqnarray}
where $M_{\alpha L}=M_{\alpha }+\frac{\pi}{c}\hat z$ and $\alpha=A,B,C$. $\Delta_{\alpha L}$ is manifestly a real number as seen from its definition. These order parameters represent charge density wave patterns of Fig.\ref{fig_order}(f) which alternate between neighboring layers. The $V_z$ term is written as:
\begin{eqnarray}
&&H_{V_z}=-\frac{V_z}{4} \sum_{j} (n_{j,\alpha}-n_{j+c\hat z,\alpha})^2 \nonumber\\
&&\ \ \ \ \ \ \ = -\frac{V_z}{4N_c}\sum_{\mathbf k\mathbf k'\mathbf q } c^\dagger_{\mathbf k+\mathbf q,\alpha}c_{\mathbf k,\alpha}(1-e^{iq_z c})c^\dagger_{\mathbf k'-\mathbf q,\alpha}c_{\mathbf k',\alpha}(1-e^{-iq_z c}).\nonumber
\end{eqnarray}
Keeping only $\mathbf q=M_{\alpha L}$, $H_{V_z}$ and its mean-field decoupling are:
\begin{eqnarray}
&&H_{V_z}= -\frac{V_z}{N_c}\sum_{\mathbf k\mathbf k'} c^\dagger_{\mathbf k+ M_{\alpha L},\alpha}c_{\mathbf k,\alpha} c^\dagger_{\mathbf k'- M_{\alpha L},\alpha}c_{\mathbf k',\alpha}, \nonumber\\
&&H_{V_z}^{MF}= -V_z \sum_{\mathbf k} \left( \Delta_{\alpha L} c^\dagger_{\mathbf k+M_{\alpha L},\alpha} c_{\mathbf k,\alpha} +h.c. \right) +V_z N_c \Delta_{\alpha L}^2.
\label{Vzdecoup}
\end{eqnarray}
For the $\lambda_z$ term in Eq.\eqref{Hcoup}, we approximate the electron density operator by the average density which is taken to be a constant. The mean field Hamiltonian also contains the three-dimensional generalization of Eq.\eqref{Vdmfu} and \eqref{Vdmfd} obtained by replacing $\Delta_{\alpha\beta},Q_{\alpha\beta}$ by $\Delta_{\alpha\beta L},Q_{\alpha\beta L}$ and the lattice distortion $u_{\alpha\beta L}$ which involves $Q_{\alpha\beta L}$. The free energy is a function of 15 variables $\mathbf \Delta=(\Delta_{AB},\Delta_{BC},\Delta_{CA},\Delta_{ABL},\Delta_{BCL},\Delta_{CAL},u_{AB},u_{BC},u_{CA},u_{ABL},$ $u_{BCL},u_{CAL},\Delta_{AL},\Delta_{BL},\Delta_{CL})$. By minimizing the free energy, we can find several types of local minima corresponding to phases that satisfy self-consistency conditions. We list some of these phases as follows:
\begin{eqnarray}
&&\Delta^+_{MMM}=(\Delta,\Delta,\Delta,0,0,0,u,u,u,0,0,0,0,0,0) \nonumber\\
&&\Delta^-_{MMM}=(-\Delta,-\Delta,-\Delta,0,0,0,-u,-u,-u,0,0,0,0,0,0) \nonumber\\
&&\Delta_{LLL}=(0,0,0,\Delta,\Delta,\Delta,0,0,0,u,u,u,\eta,\eta,\eta) \nonumber\\
&&\Delta_{MLL}=(\Delta(1 +\epsilon),0,0,0,\Delta,\Delta,u(1+\epsilon'),0,0,0,u,u,\eta,\eta,0) \nonumber\\
\label{Delform}
\end{eqnarray}
Here $\Delta$, $\eta$, $\epsilon$, $\epsilon'$ and $u$ are real numbers obtained by minimizing the free energy with $\Delta$ and $u$ being positive, and $|\epsilon|,|\epsilon'|\ll 1$. $\Delta^+_{MMM}$ ($\Delta^-_{MMM}$) represents ISD (SD) lattice distortion and charge bond order that is uniform along the z direction. $\Delta_{LLL}$ represents alternating ISD and SD patterns between neighboring layers as in Fig.\ref{fig_order3d}(a) and Fig.\ref{fig_order}(g). $\Delta_{MLL}$ represents ISD phase which is $\pi$-shifted between neighboring layers as in Fig.\ref{fig_order3d}(b) and Fig.\ref{fig_order}(h). The small $\epsilon,\epsilon'$ appearing in $\Delta_{MLL}$ is due to the broken threefold rotational symmetry. The free energy of these phases as a function of $V_d$ is shown in Fig.\ref{fig_Delta3d}(a). Here we set $V_z=0.4V_d,\lambda_z=0.4V_z, T=92K,\lambda =0.5eV/l_0,K_s =1.4eV/l_0^2$. It shows that order parameters develop as $V_d$ increases, and the $\pi$-shifted ISD phase labeled by $MLL$ has the lowest free energy. A typical value of $\Delta_{MLL}$ in Eq.\eqref{Delform} at $V_d=0.4eV$ is given by $\Delta=0.083,u/l_0=0.015,\epsilon=-0.03,\epsilon'=-0.07,\eta=-0.011$. The parameter $\Delta$ in $\Delta_{MLL}$ as a function of $V_d$ is shown in Fig.\ref{fig_Delta3d}(b). 

The ordering of these phases can be understood by examining the dependence of $F$ on $V_z$ at a fixed $V_d$, as shown in Fig.\ref{fig_Delta3d}(c). When $V_z=\lambda_z=0$, different layers are almost decoupled, then according to Sec.\ref{sec_2D} the layers with ISD order have lower free energy than those with SD order. The free energy of $\Delta_{MLL}$ and $\Delta^+_{MMM}$ are almost degenerate because they both represent ISD phases in all layers. There is a small splitting of free energy between them due to the small inter-layer hopping $t_{out}$. The $\Delta^-_{MMM}$ phase has higher free energy because it represents SD order, and the free energy of $\Delta_{LLL}$ is in the middle between $\Delta^-_{MMM}$ and $\Delta^+_{MMM}$ because it represents alternating ISD and SD orders. As $V_z$ and $\lambda_z$ increase, it favors a phase that is non-uniform along the z direction so that the effect of repulsive interaction can be reduced. Therefore, with small $V_z$ and $\lambda_z$, the free energy of $\Delta_{MLL}$ decreases and it becomes the ground state. 

If $V_z$ and $\lambda_z$ become large, $\Delta_{LLL}$ is energetically more favorable than $\Delta_{MLL}$ and becomes the ground state. This is because the charge order in neighboring layers of $\Delta_{LLL}$ are fully anti-aligned, which further reduces the repulsive interaction along the z direction. The phase diagram as a function of $V_d$ and $V_z$ is shown in Fig.\ref{fig_Delta3d}(d). Here we choose $\lambda_z=0.4V_z$. It shows the order parameters will develop as $V_d$ increases, and the ground state is the $\pi$-shifted ISD phase labeled by $\Delta_{MLL}$ when $V_z$ is small. At a large $V_d$, there is a first-order phase transition to alternating ISD and SD phase labeled by $\Delta_{LLL}$. 


\section{Discussion}

In order to take into account the electronic structure near the Fermi energy, we constructed an effective 9-band tight-binding model with V $d$ orbitals and out-of-plane Sb $p$ orbitals. We then studied the effect of intra- and inter-layer nearest-neighbor repulsive interactions on the formation of various CDW phases.
Our mean field computation has shown that various three-dimensional CDW orders arise via repulsive electron interactions and electron-lattice coupling. When the intra-layer interaction $V_d$ is strong enough, it stabilizes the CBO$^+$ phase as the ground state, which further induces ISD or SD lattice distortion depending on the sign of electron-lattice coupling. With weak inter-layer interaction $V_z$, the $\pi$-shifted (along the $c$-axis) CDW represented by $\Delta_{MLL}$ is preferred because it has lower repulsion from $V_z$ in comparison with the uniform charge order along the $c$-axis. This staggered CDW phase breaks the sixfold rotation symmetry down to twofold rotation, which is consistent with the observation in Refs.~\onlinecite{Zhao2021,Li2022s}. When $V_z$ becomes strong, the phase $\Delta_{LLL}$ with alternating SD and ISD will be the ground state because it can further reduce the repulsion from $V_z$. The emergence of these different CDW phases depends on microscopic parameters in the system and is sensitive to perturbations, which may explain the fact that these different CDW phases have all been reported in various experiments~\cite{Ritz2022,Frass2022,Luo2022s,Hu2022t}. The LCO with time-reversal symmetry breaking has not been found in our computation as the CBO order parameter is real. This is consistent with Ref.\onlinecite{Dong2022} which shows that the LCO does not occur unless there is large second-nearest-neighbor electron interaction comparable to nearest-neighbor interaction. The mechanism for time-reversal symmetry breaking is an important question that remains to be explored. Furthermore, the impact of the $\pi$-shifted CBO on superconductivity is another intriguing subject for future study.

\section{Acknowledgement}

This work is supported by the Natural Sciences and Engineering Research Council of Canada (NSERC) and the Center for Quantum Materials at the University of Toronto. H.Y.K acknowledges the support by the Canadian Institute for Advanced Research (CIFAR) and the Canada Research Chairs Program. Y.B.K. is supported by the Simons Fellowship from the Simons Foundation and the Guggenheim Fellowship from the John Simon Guggenheim Memorial Foundation. Computations were performed on the Niagara supercomputer at the SciNet HPC Consortium. SciNet is funded by: the Canada Foundation for Innovation under the auspices of Compute Canada; the Government of Ontario; Ontario Research Fund - Research Excellence; and the University of Toronto.

\appendix

\setcounter{equation}{0}
\setcounter{figure}{0}
\setcounter{table}{0}
\makeatletter
\renewcommand{\theequation}{S\arabic{equation}}
\renewcommand{\thefigure}{S\arabic{figure}}

\section{Parameters in the tight-binding model}

\begin{figure}
\centering
\includegraphics[width=3.2 in]{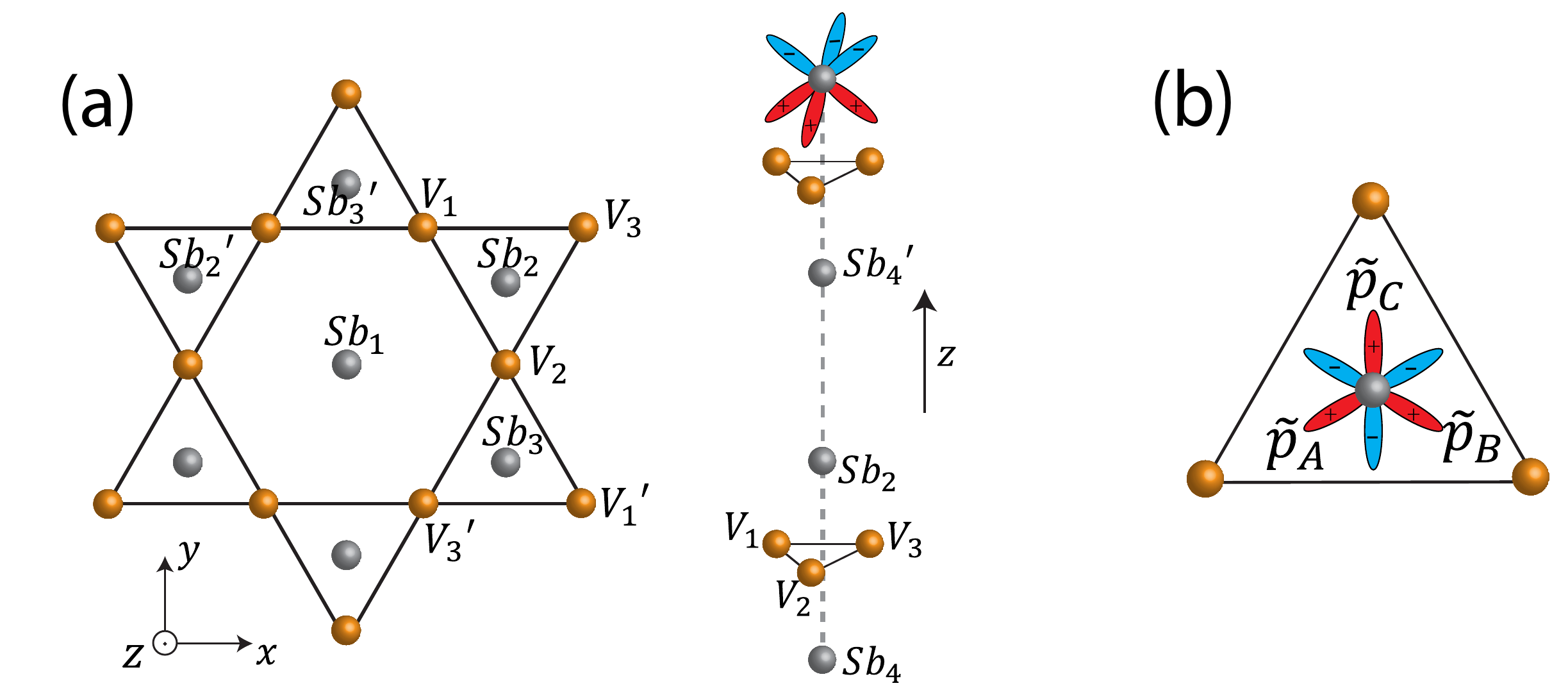}
\caption{ (a): Locations of different atomic sites to specify the tight-binding parameters. (b): The projection of three $\tilde p$ orbitals (Eq.\eqref{tildep}) to the kagome plane. The red (blue) parts represent the region of $\tilde p$ orbital with positive (negative) amplitude. These orbitals are also shown in (a). They are orthogonal and they have a finite component along the $z$ direction.  }
\label{fig_hop}
\end{figure}

Here we provide details of the tight-binding model introduced in Sec.\ref{sec_TB}. It involves five $d$ orbitals at the three vanadium sites and and three $p$ orbitals at the five antimony sites, hence there are 30 bands in total. For definiteness, we focus on CsV$_3$Sb$_5$ and obtain the hopping amplitudes and onsite potentials from DFT computation. The labeling of sites and the directions of coordinate axes are shown in Fig.\ref{fig_hop}. We choose the basis $\{d_{xz},d_{yz},d_{x^2-y^2},d_{xy},d_{z^2} \}$ and $\{p_x,p_y,p_z\}$ for $d$ and $p$ orbitals respectively. We denote the onsite potential at site $i$ by matrix $H^i$, which represents a term $c^\dagger_{i,\alpha}H^i_{\alpha\beta}c_{i,\beta}$ in the Hamiltonian where $\alpha,\beta$ represent orbitals at site $i$. The hopping amplitude between sites $i$ and $j$ is denoted by matrix $T^{i\leftarrow j}$, representing a term $c^\dagger_{i,\alpha}T^{i\leftarrow j}_{\alpha\beta}c_{j,\beta}$ in the Hamiltonian. Due to the crystalline symmetry $P6/mmm$, the three vanadium sites are related by $C_3$ rotation and their onsite potentials are not independent. The onsite potential at site $V_2$ is given by:
\bea
H^{V_2}=\begin{pmatrix}
5859.80 & 0 & 0 & 0 & 0 \\
0 & 6489.24 & 0 & 0 & 0 \\
0 & 0 & 6264.38 & 0 & -307.63 \\
0 & 0 & 0 & 5893.5 & 0 \\
0 & 0 & -307.63 & 0 & 5909.61 \\
\end{pmatrix}.
\label{hop1}
\eea
Here the basis is $\{d_{xz},d_{yz},d_{x^2-y^2},d_{xy},d_{z^2} \}$ and the unit is $meV$. The zeros in the matrix are required by symmetry. The onsite potential at $V_1$ and $V_3$ can be obtained by $C_3$ symmetry. 

For antimony sites, the four out-of-plane sites are related by symmetry. Choosing the $p$ orbital basis $\{p_x,p_y,p_z\}$, the onsite potentials at $Sb_2$ and $Sb_1$ are:
\bea
H^{Sb_2}=\begin{pmatrix}
5600.31 & 0 & 0  \\
0 & 5600.31 & 0  \\
0 & 0 & 4442.17  
\end{pmatrix}.
\eea
\bea
H^{Sb_1}=\begin{pmatrix}
4423.12 & 0 & 0  \\
0 & 4423.12 & 0  \\
0 & 0 & 5323.84  
\end{pmatrix}.
\eea
We include the NN and NNN hopping amplitudes between vanadium sites, which are given by:
\bea
T^{V_1\leftarrow V_3}=\begin{pmatrix}
341.70 & -156.09  & 0 & 0 & 0 \\
156.09 & -57.11  & 0 & 0 & 0 \\
0 & 0 & -524.12  & 112.01 & 266.03 \\
0 & 0 & -112.01 & 495.74 & -187.05 \\
0 & 0 & 266.03 & 187.05  & -112.29  \\
\end{pmatrix}.
\eea
\bea
T^{V_3\leftarrow V_1'}=\begin{pmatrix}
-34.53  & 12.74  & 0 & 0 & 0 \\
-12.74 & 9.50  & 0 & 0 & 0 \\
0 & 0 & 8.08  & -112.50  & 27.01 \\
0 & 0 & 112.50 & -24.15  & 31.04 \\
0 & 0 & 27.01 & -31.04  & 13.34  \\
\end{pmatrix}.
\eea
{From $T^{V_1\leftarrow V_3}$ we can obtain the hopping between nearest neighbor $\tilde d$ orbitals. Under the basis $\{d_{xz},d_{yz},d_{x^2-y^2},d_{xy},d_{z^2} \}$, the sixfold rotation operator is given by
\bea
C_6=\begin{pmatrix}
\frac{1}{2}  & -\frac{\sqrt{3}}{2}  & 0 & 0 & 0 \\
\frac{\sqrt{3}}{2} & \frac{1}{2}  & 0 & 0 & 0 \\
0 & 0 & -\frac{1}{2}  & -\frac{\sqrt{3}}{2}  & 0 \\
0 & 0 & \frac{\sqrt{3}}{2} & -\frac{1}{2}  & 0 \\
0 & 0 & 0 & 0  & 1  \\
\end{pmatrix}.
\label{MC6}
\eea
The $d_{xz}$ orbital can be transformed into $\tilde d$ orbital by a $\frac{\pi}{3}$ rotation at site $V_1$, or by a $-\frac{\pi}{3}$ rotation at site $V_3$. Therefore, the hopping $t_0$ between $\tilde d$ orbitals is the (1,1) element of matrix $\tilde T=(C_6)^{-1} T^{V_1\leftarrow V_3} (C_6)^{-1}$. Let the creation operator for $\tilde d$ orbital at site $j$ be $\tilde c^\dagger_j$. A direct computation shows $t_0\equiv\tilde T_{1,1}=263meV$, representing a term $t_0 \tilde c^\dagger_{V_1}\tilde c_{V_3}$ in the Hamiltonian with $t_0$ being positive.}

The hopping between vanadium and antimony sites are given as follows. It shows that the hopping between V and Sb sites are comparable or even larger than the hopping between vanadium sites.
\bea
T^{V_2\leftarrow Sb_2}=\begin{pmatrix}
-530.25  & 0  & 0  \\
0 & 555.21  & 731.17  \\
0 & 107.78 & -515.25   \\
-445.16 & 0 & 0  \\
0 & 654.61 & 65.14   \\
\end{pmatrix}.
\eea
\bea
T^{V_2\leftarrow Sb_1}=\begin{pmatrix}
0  & 0  & 637.5  \\
0 & 0 & 0  \\
-903.3 & 0 & 0   \\
0 & 776.17 & 0  \\
727.53 & 0 & 0   \\
\end{pmatrix}.
\eea
\bea
T^{V_3\leftarrow Sb_3}=\begin{pmatrix}
-18.73   & -107.17   & 33.27  \\
-34.99  & -92.29  & -43.12   \\
-15.01  & 4.36 & -29.04    \\
-56.00  & 13.31 & 47.50   \\
22.93 & 62.87 & 5.67   \\
\end{pmatrix}.
\eea
The hopping between antimony sites are:
\bea
T^{Sb_2\leftarrow Sb_3}=\begin{pmatrix}
-485.53 & 0 & 0  \\
0 & 1638.32 & 89.63  \\
0 & -89.63 & -420.64  
\end{pmatrix}.
\eea
\bea
T^{Sb_1\leftarrow Sb_3'}=\begin{pmatrix}
47.94 & 0 & 0  \\
0 & 487.18 & 572.10  \\
0 & 350.30 & 197.82  
\end{pmatrix}.
\eea
\bea
T^{Sb_2'\leftarrow Sb_2}=\begin{pmatrix}
-233.02 & 15.07 & -24.81  \\
-15.07  & -35.31 & -3.29   \\
24.81 & -3.29  & 36.80  
\end{pmatrix}.
\eea
\bea
T^{Sb_2\leftarrow Sb_4}=\begin{pmatrix}
12.75 & 0 & 0  \\
0 & 12.75 & 0  \\
0 & 0 & 582.20  
\end{pmatrix}.
\eea
\bea
T^{Sb_2\leftarrow Sb_4'}=\begin{pmatrix}
-11.12 & 0 & 0  \\
0 & -11.12 & 0  \\
0 & 0 & 154.94  
\end{pmatrix}.
\label{hopf}
\eea
All these matrices describe distinct hopping processes that are unrelated by symmetry, and the full model is generated by acting elements of $P6/mmm$ to each hopping process. Among these hopping processes, only $T^{Sb_2\leftarrow Sb_4'}$ connects different unit cells separated along the $c$-axis. This is the $t_{out}$ term mentioned in the main text. If $T^{Sb_2\leftarrow Sb_4'}$ is set to zero, the band dispersion will be uniform along $k_z$. 

We can simplify this tight-binding model by focusing on the van-Hove singularities at $M$ point which mainly consists of $d_{xz}$ and $d_{yz}$ orbitals at vanadium sites. By choosing a suitable linear combination of $d_{xz}$ and $d_{yz}$ orbitals at each vanadium sites as indicated by the red orbitals in Fig.\ref{fig_crystal}(a), the orbitals $\tilde d_\alpha$ ($\alpha=A,B,C$) contribute more than 80\% weight of the wave function at the vH points. The $\tilde d_\alpha$ orbitals are given by:
\bea
\tilde d_A&=&-\frac{1}{2}d_{A,xz}+\frac{\sqrt{3}}{2}d_{A,yz}\nonumber\\
\tilde d_B&=&-\frac{1}{2}d_{B,xz}-\frac{\sqrt{3}}{2}d_{B,yz} \nonumber\\
\tilde d_C&=&d_{C,xz}
\eea
We keep only one $\tilde d_\alpha$ orbital out of the five $d$ orbitals at each vanadium sites. We also neglect the three $p$ orbitals at the in-plane antimony site, because these orbitals mainly contribute to the $\Gamma$ point and have little contribution to the vH point. This reduces the 30-band model to a 15-band model with three $\tilde d$ orbitals and twelve $p$ orbitals at the four out-of-plane Sb sites. 

We can further simplify this model by finding a more symmetric basis for the $p$ orbitals. For each Sb site, instead of using the global $x,y,z$ directions for the $p$ orbitals, we can use the rotated basis in which the projections of the three orthogonal $p$ orbitals in the kagome plane are symmetrically aligned with the neighboring vanadium sites, as in Fig.\ref{fig_hop}(b). Denote these $p$ orbitals by $\tilde p_{\alpha}$ whit $\alpha=A,B,C$, which are given by:
\bea
\tilde p_{A}&=&-\frac{1}{\sqrt{2}}p_x-\frac{1}{\sqrt{6}}p_y-\frac{1}{\sqrt{3}}p_z, \nonumber\\
\tilde p_{B}&=&\frac{1}{\sqrt{2}}p_x-\frac{1}{\sqrt{6}}p_y-\frac{1}{\sqrt{3}}p_z, \nonumber\\
\tilde p_{C}&=&\sqrt{\frac{2}{3}}p_y-\frac{1}{\sqrt{3}}p_z.
\label{tildep}
\eea
The $\tilde p$ orbitals at the other out-of-plane Sb sites are generated by transforming Eq.\eqref{tildep} by reflection symmetry $M_z$ across the kagome plane and twofold rotation $C_{2z}$ around vanadium sites. Noticing that the $\tilde p$ orbitals at sites below and above the kagome plane can be superimposed to yield orbitals with positive or negative eigenvalues of the in-plane mirror symmetry $M_z$. The wave function at the vH point must be an eigenstate of $M_z$. Since the $\tilde d$ orbitals have negative eigenvalues of $M_z$, then among the twelve $\tilde p$ orbitals only the six orbitals with negative $M_z$ eigenvalue can contribute to the wave function at the vH point. Denote these six orbitals by $\tilde p_{\sigma\alpha}^-$ with $\alpha=A,B,C$ and $\sigma=1,2$ representing the two sublattices of Sb sites above the kagome plane, e.g., $\tilde p_{1\alpha}^-$ is the superposition of $\tilde p_\alpha$ at Sb$_3$ and its mirror image with $M_z$ eigenvalue $-1$, and $\tilde p_{2\alpha}^-$ is the superposition of $\tilde p_\alpha$ at Sb$_2$ and Sb$_4$ with $M_z$ eigenvalue $-1$. The similar superposition with $M_z$ eigenvalue $+1$ is denoted as $\tilde p_{\sigma\alpha}^+$. Combining the three $\tilde d_\alpha$ orbitals and the six $\tilde p_{\sigma\alpha}^-$ orbitals, we obtain the basis for the 9-band model mentioned in the main text. 

{To obtain the matrix for this 9-band model, we first need to perform a unitary transformation to the 30-band model. The original 30-band model $H_{TB}(\mathbf k)$ can be obtained from the DFT parameters in Eqs.\eqref{hop1}-\eqref{hopf}. The ordering of the orbitals in $H_{TB}(\mathbf k)$ are chosen as $ \{d_{V_3'},d_{V_1'},d_{V_2},p_{\mathrm{Sb}_3},p_{\mathrm{Sb}_5},p_{\mathrm{Sb}_2},p_{\mathrm{Sb}_4},p_{\mathrm{Sb}_1} \}$ according to Fig.\ref{fig_hop}(a) with Sb$_5$ being the mirror image of Sb$_3$ below the kagome plane. Here $d_{V_i}$ represents the five orbitals $d_{xz},d_{yz},d_{x^2-y^2},d_{xy},d_{z^2}$ at site $V_i$ and $p_{\mathrm{Sb}_i}$ represents the three orbitals $p_x,p_y,p_z$ at site Sb$_i$. The $x,y,z$ directions in these orbitals refer to the global coordinate directions in Fig.\ref{fig_hop}(a). We apply a unitary transformation $\mathcal U$ to rotate the basis such that it involves the local $\tilde d_\alpha$ and $\tilde p_{1\alpha}^\pm$ orbitals:
\be
\tilde H_{TB}(\mathbf k)=\mathcal{U} H_{TB}(\mathbf k) \mathcal{U}^\dagger,\ \ \mathcal U=\mathcal U^a \mathcal U^b
\label{rotH}
\ee
The $30\times 30$ matrices $\mathcal U^a$ and $\mathcal U^b$ are given as follows. Define matrices:
\bea
U_1=\begin{pmatrix}
-\frac{1}{\sqrt{2}} & -\frac{1}{\sqrt{6}} & -\frac{1}{\sqrt{3}}  \\
\frac{1}{\sqrt{2}} & -\frac{1}{\sqrt{6}} & -\frac{1}{\sqrt{3}} \\
0 & \sqrt{\frac{2}{3}} & -\frac{1}{\sqrt{3}}  
\end{pmatrix},\ U_2=\begin{pmatrix}
\frac{1}{\sqrt{2}} & \frac{1}{\sqrt{2}}   \\
\frac{1}{\sqrt{2}} & -\frac{1}{\sqrt{2}} 
\end{pmatrix}.
\eea
Here $U_1$ comes from Eq.\eqref{tildep} and $U_2$ gives the linear superposition between mirror-related sites above and blow the kagome plane. The threefold rotation matrix is $C_3\equiv (C_6)^2$ with $C_6$ defined in Eq.\eqref{MC6}. let $\mathbb I_n$ be the $n\times n$ identity matrix. Denote $M_{m\sim n}$ as the diagonal block of a matrix $M$ made of the $m$-th to $n$-th rows and columns. Then $\mathcal {U}^a$ is defined by $\mathcal{U}^a_{1\sim 5}=C_3^\dagger$, $\mathcal{U}^a_{6\sim 10}=C_3$, $\mathcal{U}^a_{11\sim 15}=\mathbb I_5$, $\mathcal{U}^a_{16\sim 27}=\mathbb{I}_2\otimes U_2\otimes U_1$, $\mathcal{U}^a_{28\sim 30}=\mathbb I_3$. Here $\otimes$ is the direct product. $\mathcal U^b$ is defined by replacing the 21-th and 24-th diagonal elements of $\mathbb I_{30}$ by $-1$, i.e., $\mathcal U^b_{21,21}=\mathcal U^b_{24,24}=-1$ and the other elements of $\mathcal U^b$ is the same as $\mathbb I_{30}$. The matrix $\mathcal U^b$ takes into account the different orientations of $\tilde p$ orbitals at sites above and below the kagome plane because $p_z$ flips sign under mirror symmetry $M_z$, and the matrix $\mathcal U^a$ implements the rotation to the new local basis involving $\tilde d_\alpha$ and $\tilde p_{\sigma\alpha}^\pm$ orbitals. }

{The orbitals in the 9-band model are $\{\tilde \Phi\}=\{\tilde d_A,\tilde d_B,\tilde d_C,\tilde p_{1A}^-,\tilde p_{1B}^-,\tilde p_{1C}^-,\tilde p_{2A}^-,\tilde p_{2B}^-,\tilde p_{2C}^- \}$, which correspond respectively to the 1,6,11,19,20,21,25,26,27-th orbitals of the rotated Hamiltonian $\tilde H_{TB}(\mathbf k)$ obtained via Eq.\eqref{rotH}. By performing another unitary transformation such that these nine orbitals become the first nine orbitals, we arrive at a $30\times 30$ matrix $\tilde H_{TB}'(\mathbf k)$ in which the first nine orbitals are given by $\{\tilde \Phi\}$. Then the matrix of the 9-band model discussed in the main text is obtained by the first $9\times 9$ block of $\tilde H_{TB}'(\mathbf k)$.}

The dispersion of this 9-band model is in Fig.\ref{fig_crystal}(d). It captures the van-Hove singularities vH1 and vH2. Using the basis $\{\tilde d_A,\tilde d_B,\tilde d_C,\tilde p_{1A}^-,\tilde p_{1B}^-,\tilde p_{1C}^-,\tilde p_{2A}^-,\tilde p_{2B}^-,\tilde p_{2C}^- \}$, the components of the squared wave function at vH1 at momentum $M_c$ is $(0,0,a,b,b,0,b,b,0)$ with $a=0.92,b=0.02$. Therefore vH1 is dominated by $\tilde d$ orbitals. The squared wave function at vH2 at momentum $M_c$ is $(a',a',0,0,0,b',0,0,b')$ with $a=0.27,b=0.23$. Therefore vH2 is a mixture of $d$ and $p$ orbitals. The simple form of wave function is due to our choice of symmetric basis, which is helpful for modeling the van-Hove singularities.


%

\end{document}